\parindent=0pt

\def\square{\,\hbox{\vrule\vbox{\hrule\phantom{N}\hrule}\vrule}\,}

\centerline{\bf OLD AND NEW RESULTS FOR  SUPERENERGY TENSORS } 

\centerline{\bf  
FROM} 

\centerline{\bf DIMENSIONALLY DEPENDENT TENSOR IDENTITIES}

\

\

\centerline{\bf S. Brian
Edgar and Ola Wingbrant }

\centerline{Department of Mathematics, }

\centerline{Link\"{o}pings universitet,}

\centerline{Link\"{o}ping,}

\centerline{Sweden S-581 83.}
\smallskip
\centerline{ email: bredg@mai.liu.se,\ olawi214@student.liu.se} 

\

\


\

\

{\bf Abstract.}

It is known that some results for spinors, and in particular for superenergy spinors,  are much less transparent
and require a lot more effort to establish, when considered from the tensor viewpoint. In this paper we
demonstrate how the use of dimensionally dependent tensor identities enables us to derive a number of  $4$-dimensional
identities by straightforward tensor methods in a signature independent manner.  In particular, we  consider the quadratic
identity for the Bel-Robinson tensor
${\cal T}_{abcx}{\cal T}^{abcy} =
\delta_x^y \ {\cal T}_{abcd}{\cal T}^{abcd}/4$ and also  the new conservation laws for the Chevreton tensor, both of which
have been obtained by spinor means; both of these results are rederived by   {\it  tensor} means for $4$-dimensional spaces
of any signature,  using dimensionally dependent
 identities, and also we  are able to conclude that there are no {\it direct} higher dimensional analogues. In addition we
demonstrate a simple way to show  non-existense of such identities  via counter examples; in particular we show that there  is
no  non-trivial  Bel tensor analogue of this simple Bel-Robinson tensor quadratic identity. On the other hand, as a sample of
the power of generalising  dimensionally dependent tensor identities from four to  higher dimensions,  we show that the
symmetry structure, trace-free and divergence-free nature of the four dimensional   Bel-Robinson tensor does have an analogue
for a class of tensors in higher dimensions.

\

\

{\bf 1 Introduction.}

Investigations connected with the Bel-Robinson tensor [1] in four dimensions are usually much simpler and more
efficient when carried out in spinor formalism [2]. Senovilla [3] has demonstrated that a much larger class of
tensors --- {\it superenergy tensors} --- share most of the desirable properties of the Bel-Robinson tensor, and 
Bergqvist [4] has shown how {\it superenergy  spinors}  give a simpler and more efficient presentation of  certain
aspects of these superenergy tensors. Recently, Bergqvist, Eriksson and Senovilla [5] have obtained  new
conservation laws for the electromagnetic field using superenergy spinor considerations, and emphasised that  the
proof of this result is far from obvious from the tensor point of view.

It is reassuring to know that certain important but perhaps unexpected properties ---  disguised in the
complexities of the tensor formalism --- become more transparent in the spinor formalism; but the parallel
and more transparent spinor investigations are restricted to
$4$-dimensional spacetimes with Lorentz signature, and so this assistance is
not available  in higher dimensions nor in four dimensional spaces with other signatures.

Deser [6] has emphasised the significance of the Bel-Robinson tensor in
higher dimensions, and one of the important features of Senovilla's method of construction of superenergy
tensors [3] is that it is applicable to arbitrary fields {\it in any dimension}; and so, for higher dimensions, it becomes an
obvious concern whether there could be unexpected  properties for superenergy
tensors --- disguised in the even deeper complexities of   tensor formalism  in higher dimensions ---
analogous to those  properties revealed by spinor formalism in four dimensions.

Deeper investigations into the  interaction between  dimension and tensor identities have been instrumental
in illustrating the uniqueness of some of the Bel-Robinson tensor's properties in four dimensions [6],  
explaining the collapse of some Riemann scalar invariants in four dimensions [7],  resolving apparent disparities
between the spinor and tensors versions of the wave equations for the Weyl tensor and Lanczos potential
[8,9,10,11] respectively.  Moreover, in higher dimensions, worries
 concerning counterterms in Lagrangians [12,13] have been dispelled, and the Bel-Robinson tensor has been
shown to be fully symmetric in five dimensions (as well as in four dimensions) [3].

Much earlier, Lovelock [14] had pointed out that a number of apparently unrelated results were all really
consequences of a class of identities which he christened {\it dimensionally dependent identities} ---
identities which are a trivial, but subtle, consequence of dimension alone. Recently Edgar and H\"oglund [15]
have generalised Lovelock's results,  and  demonstrated  that the underlying principle in all of these
investigations in [6 - 13], and some new ones,  was the explicit exploitation   of dimensionally dependent
identities. Furthermore,    in 
algebraic Rainich theory,  Bergqvist and H\"oglund [16] have exploited these ideas further, and obtained results
in {\it five} dimensions involving {\it cubic} terms in the energy momentum tensor --- motivated by  the familiar
results in {\it four} dimensions involving {\it quadratic} terms, [17]; while Edgar and H\"oglund [18] have
demonstrated the crucial role that dimensionally dependent identities play in the existence of the Lanczos
potential for the Weyl tensor in different dimensions.

 Deser [19] has applied
the  adjective  'ubiquitous' to the Bel-Robinson tensor, and  this description is  equally appropriate to
dimensionally dependent identities  as can be seen by the wide range of the
investigations in [6 - 13],  the  applications given by  Lovelock [14], and the more recent applications in
[15,16,18].     In fact, Deser  has argued elsewhere [6], that two identities (which are
examples of what we call dimensionally dependent identities in four dimensions) 
 are in a sense implicit in the  familiar definitions of the Bel-Robinson tensor in four dimensions.

The purpose of this present paper is    
 to emphasise the subtle interaction between dimension and tensor identities, and illustrate the important role
which can be played by {\it fundamental dimensionally dependent identities} in investigations where
tensor identities are important, in particular involving superenergy tensors. (We shall refer to the most
fundamental dimensionally dependent identities as 'fddis', and to any identities constructed from these as
'ddis'). Our overall aim is to examine  useful and significant properties in four
dimensions --- usually originating as  spinor identities ---  and identify the kernel $4$-dimensional fddi;
then we will use the  higher dimensional analogues of the kernel fddi to try and establish analogous results in
higher dimensions. This will involve two different stages of investigation:

\smallskip

{\it Step 1. }
The first step is to establish {\it
$4$-dimensional signature-independent tensor versions and proofs} of  interesting  spinor identities and/or
reconcile apparent discrepencies between spinor and tensor results. We stress the need for {\it
signature-independent} proofs for the following reason: results obtained using spinors strictly can  claim to be  valid  only
in $4$-dimensional spacetimes {\it with Lorentz signature}. Of course there are results in such spaces which have no
counterpart in other signatures (e.g., results concerning principle null directions of the Weyl tensor), but we encounter an
uncertain  situation  when we consider results which can be stated in tensors with no apparent reference to
signature, but which were derived in spinors, or derived in tensors but using features which are signature
dependent\footnote{${}^{\dag}$}{When some results, which had been obtained by spinor means for the Weyl spinor
[12] and Lanczos spinor [20],  proved  difficult to reproduce by tensor analysis there was  speculation
in
[12] and  [20] respectively that these  results could be obtained {\it only} by spinor
means. However, in these particular cases, signature independent versions were obtained by use of
$4$-dimensional tensor fddis. in [13,15] and [11] respectively.}. The familiar identity for the Bel-Robinson
tensor
$${\cal T}_{abcx}{\cal T}^{abcy} =
\delta^y_x \ {\cal T}_{abcd}{\cal T}^{abcd}/4 \eqno(1)$$  in four dimensions is an important example of such a
situation;  we would wish to understand whether  dimension and signature have crucial roles in this
result\footnote{${}^{\ddag}$}{Identity (1)  is quoted in [21] with a reference to a
proof by Debever in [22]; this proof is a lot more complicated than the spinor proof in [2]. Moreover, Debever's
proof is also explicitly for a Lorentz spacetime; he makes  use  of the principle null directions, and
it is not easy to see how this proof could  be generalised to other signatures of 
$4$-dimensional spaces, or to higher dimensions.  Hence the identity  has  really only been proven in [2] and [22]
for $4$-dimensional spacetimes {\it with Lorentz signature}, and strictly its applicability to other signatures
has not been confirmed in those proofs. }.

In spinor calculations
the dimension four is inbuilt into the formalism;  in tensor calculations  a non-arbitrary dimension such as four
has to be put in explicitly 'by hand'. However, it is not always sufficient just to substitute $n=4$ in explicit
calculations; in some cases the substitution needed is more subtle --- it is achieved
by the  use of one or more fddis, but it is  clear that there is no direct 
spinor analogue of a $4$-dimensional fddi --- the spinor version is trivially zero.

So, for example,  the 
Lanczos spinor potential for the Weyl spinor 
$L_{ABCD'}=L_{(ABC)D'}$,  in Ricci flat spaces, was found by Illge [10] to satisfy the very simple equation
$$\square L_{ABCD'}=0
\eqno(2)$$
while the corresponding tensor equation for the Lanczos tensor
$L_{abc}=L_{[ab]c}, \ L^c{}_{ac}=0= L_{[abc]}$  is calculated to be [9,11]
$$
\eqalign{
\nabla^2 L_{abc}
  +{2(n-4)\over n-2}L_{[a}{}^{d}{}_{|c;d|b]}
=2L_{[b}{}^{ed}C_{a]dec}
  -{1\over 2}C_{deab}L^{de}{}_{c}
  +{4\over n-2}g_{c[a}C_{b]fed}L^{fed}
  }
\eqno(3)$$
where $C_{abcd}$ is the Weyl tensor. In four dimensions, obviously it cannot be  sufficient simply to substitute 
$n=4$, since we know from the spinor version that  the right hand side must disappear completely in four
dimensions. But if we consider the
$4$-dimensional fddi
$C_{[ab}{}^{[cd}
\delta_{f]}^{e]}\equiv  0$ [14,15] (quoted in Lemma 3 at the end of this section), we find that,  when contracted
with 
$L_{de}{}^f$, we obtain the ddi,
$$
\eqalign{
2L_{[b}{}^{de}C_{a]edc}-{1\over2} L^{de}{}_c 
C_{deab}
+  2L^{def}g_{c[a}C_{b]def} \equiv 0}
\eqno(4)$$
which ensures that the whole of the right hand side of (3) disappears in four dimensions [10]. 

\smallskip

{\it Step 2. } Once the $4$-dimensional version is fully understood and the  $4$-dimensional kernel fddi
obtained, the second step will be to determine what generalisations are possible using the higher dimensional
counterpart fddis of the  $4$-dimensional kernel fddi. Occasionally these generalisations can be quite
straightforward (e.g., discovering that a $4$-dimensional result is also valid in  five dimensions [3,15]);  or
more complicated involving a restructuring of the $4$-dimensional result in higher dimensions (e.g., finding
a $5$-dimensional result involving triple products of Maxwell tensors as a generalisation of a $4$-dimensional
result involving double products of Maxwell tensors [16]).

\medskip

The remainder of the paper is organised as follows. 
In Section 2 we deduce four different spinor identities  which are  special cases of one very
simple general spinor identity; but in Section 3 we find that the
$4$-dimensional signature-independent tensor version of each of these identities requires a very different {\it
tensor} proof --- some of which are very complicated --- although the unifying characteristic in all is the use
of ddis.

In Section 4 we show that  the Bel superenergy tensor does not satisfy
the same simple  identity (1) in four dimensions, and in Section 5 we show that the Bel-Robinson tensor
does not satisfy any analogous identity to (1) in {\it five} dimensions. In Section 7 we  rederive the new
conservation laws for electromagnetic theory [5] by using a number of
$4$-dimensional tensor ddis; these $4$-dimensional tensor ddis cannot be replaced {\it directly} with higher dimensional ddis,
and so there is no {\it direct} higher dimensional analogue of this law.

As noted above, the second step in such investigations is
to attempt generalisation of
$4$-dimensional results  to  higher dinensions once we have identified the kernel tensor fddi in four
dimensions.
In Section 6 we
give the
$4$-dimensional  tensor counterparts to two trivial spinor results involving the symmetry properties of the
Bel-Robinson and Lanczos superenergy spinors: these results both involve $4$-dimensional fddis, and in the case
of the Bel-Robinson superenergy tensor, by means of the analogous
$5$-dimensional fddi we show that exactly the same result is true in five dimensions.

A more ambitious generalisation is proposed in Section 8. We illustrate this approach by considering
a  superenergy tensor which is a natural generalisation of the Bel-Robinson tensor, and show that it shares
its attractive properties of full index symmetry and zero divergence in {\it seven  and lower} dimensions, as
well as being trace-free in {\it six} dimensions.   A summary is given, and future
developments are proposed in Section 9. 

\smallskip

It will be useful to have for reference a number of   lemmas which are simply
tensor ddis in four dimensions. Many familiar identities ostensibly involve the Weyl or Riemann
tensors directly or indirectly, but on closer inspection have a
more general character being simply algebraic, involving 'candidates'\footnote{${}^{\dag}$}{A
'candidate' of a tensor such as the Riemann or Weyl tensor is a tensor
with the same index and trace properties, but sharing no other
properties, such as differential properties; we shall designate such
candidates with the symbol \ $\hat{}$\ . So, for example a 'Weyl
candidate tensor' $\hat C_{abcd}$ is defined by the properties
$\hat C_{abcd}=\hat C_{[ab]cd}=\hat C_{ab[cd]}, \ \hat C_{a[bcd]}=0, \
\hat C^a{}_{bca}=0.$ We shall follow the usual notation [2] with
$R_{abcd}, C_{abcd}, S_{ab}, R$ for Riemann, Weyl, trace-free Ricci tensors and
Ricci scalar respectively; their 'candidates' will be respectively $\hat R_{abcd},
\hat C_{abcd}, \hat S_{ab}, \hat R $.  We shall also follow the usual notation for the  Weyl and Ricci spinors
respectively
$\Psi_{ABCD},
\Phi_{ABC'D'}$, and  scalar $ \Lambda (= R/24 )$; their 'candidates' will be respectively,
$\hat\Psi_{ABCD},
\hat\Phi_{ABC'D'}$, $\hat \Lambda $. In addition we will use the Lanczos spinor $L_{ABCD'}=L_{(ABC)D'}$, and
Lanczos tensor ${L}_{abc}=L_{[ab]c}, \  L^c{}_{ac}=0=  L_{[abc]}$ [9,11] with corresponding 'candidates' $\hat
{L}_{ABCD}$ and
 $\hat L_{abc}$.} for Weyl, Riemann, Lanczos 
 or other tensors. In this paper, we shall give the results for the more general 'candidates'
 where appropriate.

The following lemmas can be found in [15], or can be deduced from results there. The first three of these
lemmas
are fddis obtained by skew symmetrising over {\it five} indices in
$4$-dimensional space (and {\it six} indices in
$5$-dimensional space), and  exploiting the fact that the appropriate tensors are
trace-free;  Lemma 4 gives ddis deduced from Lemma 3, but as can be seen from the details in [23],
although (8a) is well-known, quite a lot of work is involved in obtaining (8b,c,d).

{\bf Lemma 1.} In four dimensions,  a 2-tensor $A_{ab}$  
satisfies
$$A_{[a}{}^{a}  A_b{}^b A_c{}^c A_d{}^d\delta_{e]}^{f} \equiv 0 \ , \eqno(5)$$
which is equivalent to the
Cayley-Hamilton Theorem for the $4 \times 4$ matrix 
$A_a{}^b$ when written out term by term. 

(We shall be concerned with two special cases from this class of tensors: trace-free Ricci candidates $\hat
S_{ab}=\hat
S_{(ab)} $ with $\hat
S^a{}_{a}=0 $, and Maxwell tensors $ F_{ab}=  F_{[ab]}$.)

{\bf Lemma 2.}   A Lanczos candidate $\hat {L}_{abc}$  with properties
$\hat{L}_{abc}=\hat  L_{[ab]c},
\ \hat L^c{}_{ac}=0= \hat L_{[abc]}$, satisfies
$$\hat{L}_{[ab}{}^{[e}\ \delta_{cd]}^{fg]} \equiv 0 \qquad \hbox{in four dimensions} ,\eqno(6a)$$
$$\hat{L}_{[ab}{}^{[f}\ \delta_{cde]}^{ghi]} \equiv 0 \qquad \hbox{in five dimensions} .\eqno(6b)$$

{\bf Lemma 3.} A Weyl candidate $\hat C_{ab}{}^{cd}$  satisfies
$$\hat C_{[ab}{}^{[de}\ \delta_{c]}^{f]} \equiv 0 \qquad \hbox{in four dimensions}, \eqno(7a)$$
$$\hat C_{[ab}{}^{[ef}\ \delta_{cd]}^{gh]} \equiv 0 \qquad \hbox{in five dimensions}. \eqno(7b)$$

{\bf Lemma 4.} In four dimensions, a Weyl candidate  $\hat C_{ab}{}^{cd}$ satisfies
$$\hbox {(a)} \qquad\hat C_{abcx}\hat {C}^{abcy}\equiv \delta^y_x \ \hat C_{abcd} \hat{C}^{abcd}/4 \ ,\eqno(8a)$$
$$\hbox {(b)} \qquad \qquad\hat C^{yb}{}_{cd}\hat C^{de}{}_{bf}\hat C^{fc}{}_{ex}  \equiv \delta^y_x \
 \hat C^{ab}{}_{cd}\hat C^{de}{}_{bf}\hat C^{fc}{}_{ea}
/4 \ , \qquad \ \eqno(8b)$$
$$\hbox {(c)} \qquad\hat C^{yb}{}_{cd}\hat C^{de}{}_{bf}\hat C^{fg}{}_{eh} \hat
C^{hc}{}_{gx} \equiv \delta^y_x \
 \hat C^{ab}{}_{cd}\hat C^{de}{}_{bf}\hat C^{fg}{}_{eh}
\hat C^{hc}{}_{ga}/4 \ ,\eqno(8c)$$
$$\hbox {(c)} \qquad\hat C^{yb}{}_{cd}\hat C^{de}{}_{bf}\hat C^{fgc}{}_{h} \hat C^{h}{}_{egx}\equiv \delta^y_x \
\hat C^{ab}{}_{cd}\hat C^{de}{}_{bf}\hat C^{fgc}{}_{h} \hat C^{h}{}_{ega}/4  \ .\eqno(8d)$$

It is obvious from the constructions that  these identities for four/five dimensions are valid in four/five {\it
and lower} dimensions. So, for instance (6b) is also valid in four dimensions, but the only
non-trivial information is in its trace in four dimensions, which is equivalent to (6a); on the otherhand, in
five dimensions, the trace of the left hand side of (6b) is identically zero, giving a trivial result.  Of
course no Weyl candidates exist in dimensions less than four, but for the other tensors, these  lemmas are
nontrivial in lower dimensions. However, we shall not be concerned with dimensions less than four in this paper.
Although stated for Lanczos and Weyl candidates (which is all we require in this paper), a number of these results are
valid for more general tensors; in particular the antisymmetry property $\hat L_{[abc]}=0$ can be relaxed in Lemma 2, and the
antisymmetry
$\hat C_{a[bcd]}=0$ can be relaxed in Lemma 3 and Lemma 4(a,b).

\

\

{\bf 2. Simple Spinor Identities}.

We  begin with the following spinor result which generalises Penrose's
original derivation [2] for Bel-Robinson tensors: 

{\bf Theorem 1.}  A spinor which factorises according to $\top_{{\cal
A}X{\cal A}'X'} = 4 V_{{\cal S}X}\bar{V}_{{\cal S}'X'}$ satisfies
$$\top_{{\cal S}X{\cal S}'X'}\top^{{\cal S}Y{\cal S}'Y'} =
\epsilon_X{}^Y  \epsilon_{X'}{}^{Y'} \top_{{\cal S}A{\cal S}'A'}\top^{{\cal S}A{\cal S}'A'}/4 \eqno(9)$$
where ${\cal S}, {\cal S}'$ each represent an {\it odd} number of
spinor indices.

{\bf Proof.} 
$$V_{{\cal S}X} V^{{\cal S}Y} = V_{{\cal S}}{}^Y V^{{\cal S}}{}_X +
\epsilon_X{}^Y V_{{\cal S}A} V^{{\cal S}A} = -V^{{\cal S}}{}^Y V_{{\cal S}}{}_X +
\epsilon_X{}^Y V_{{\cal S}A} V^{{\cal S}A}
$$
with the negative sign arising by 'see-sawing' the odd number of indices in
${\cal S}$. Hence
$$V_{{\cal S}X} V^{{\cal S}Y} =
\epsilon_X{}^Y V_{{\cal S}A} V^{{\cal S}A}/2\eqno(10)
$$
Multiplying by the complex conjugate 
$$\bar{V}_{{\cal S'}X'} \bar{V}^{{\cal S'}Y'} V_{{\cal S}X} V^{{\cal S}Y}
=
\epsilon_X{}^Y V_{{\cal S}A} V^{{\cal S}A} \epsilon_{X'}{}^{Y'}
\bar{V}_{{\cal S'}A'} \bar{V}^{{\cal S'}{A'} }/4
$$
and substituting for $\top_{{\cal S}X{\cal S}'X'}$, $\top_{{\cal S}Y{\cal S}'Y'}$ gives the result.  \hfill $\diamondsuit$

From Theorem 1 we see that the types of indices in the collection of
indices represented by ${\cal S}$ do not matter; only the fact that
there is an odd number. In this paper we shall be concentrating on
$4$-index tensors
${\cal T}_{abcd}$ equivalently $ {\cal T}_{ABCDA'B'C'D'}$; and in particular from [4],

$\bullet$   the superenergy spinor of the 
Weyl (candidate) spinor $\hat \Psi_{ABCD}$ (i.e., the
Bel-Robinson  superenergy spinor) is given by
$${\cal
T}[\hat\Psi]_{ABCDA'B'C'D'}  = 4 \hat{\Psi}_{ABCD}\bar
{\hat\Psi}_{A'B'C'D'} \eqno(11a)$$

$\bullet$ the superenergy spinor of
the Ricci (candidate) spinor $\hat
{\Phi}_{ABC'D'}$ is given by
$$
 {\cal T}[\hat\Phi]_{ABCDA'B'C'D'}  = 4 \hat \Phi_{ABC'D'}\bar{\hat
{\Phi}}_{CDA'B'} 
\eqno(11b)$$

$\bullet$  the superenergy spinor of the  Ricci (candidate) scalar $\hat
\Lambda$ is given by 
$$
 {\cal T}[\hat\Lambda]_{ABCDA'B'C'D'}  = 4 {\hat \Lambda}^2 (\epsilon _{AC}\epsilon _{BD}+\epsilon
_{AD}\epsilon _{BC})(\epsilon _{A'C'}\epsilon _{B'D'}+\epsilon _{A'D'}\epsilon
_{B'C'})
  \eqno(11c)$$

$\bullet$  the superenergy spinor of the  Weyl-Ricci scalar (candidate) spinor $\hat
{\chi}_{ABCD}$ is given by 
$$ \eqalign{{\cal T}[\hat\chi]_{ABCDA'B'C'D'} & = 4\hat\chi_{ABCD}\bar{\hat\chi}_{A'B'C'D'} \cr & =   4 \Bigl( \hat
\Psi_{ABCD}+
\hat\Lambda(\epsilon _{AC}\epsilon _{BD}+\epsilon _{AD}\epsilon _{BC})\Bigr)\Bigl(\bar{
{\hat\Psi}}_{A'B'C'D'}+\hat \Lambda(\epsilon _{A'C'}\epsilon _{B'D'}+\epsilon _{A'D'}\epsilon
_{B'C'})\Bigr)\cr & = {\cal T}[\hat \Psi] + {\cal T}[\hat \Lambda]\cr & \qquad + 4 \hat \Lambda\Bigl(\bar{
{\hat\Psi}}_{A'B'C'D'}(\epsilon _{AC}\epsilon _{BD}+\epsilon _{AD}\epsilon _{BC})+ \hat
\Psi_{ABCD}(\epsilon _{A'C'}\epsilon _{B'D'}+\epsilon _{A'D'}\epsilon
_{B'C'})\Bigr)
\ .}\eqno(11d)$$

Theorem 1 immediately specialises to,

{\bf Theorem 2.} 
The four superenergy spinors in (11) 
\ ${\cal
T}[\hat\Psi],\  
 {\cal T}[\hat\Phi], \ {\cal T}[\hat\Lambda], \ {\cal T}[\hat\chi]$ \ all obey the
identity
$$
{\cal T}_{ABCXA'B'C'X'} {\cal T}^{ABCYA'B'C'Y'}  = \epsilon_X{}^Y
\epsilon_{X'}{}^{Y'}{\cal T}_{ABCDA'B'C'D'} {\cal T}^{ABCDA'B'C'D'}/4 \ .
\eqno(12)$$
\hfill $\diamondsuit$

The simplicity of the above theorems is due to the fact that the superenergy spinors were
simple direct products involving a spinor  times its conjugate; it should be noted that even more
general identities could have been  obtained for these four spinors, as well as for more general spinors with
the same simple product structure --- not just from the point of view of relaxing the index symmetries, but also
from freeing more indices, [2]. 

However the Bel superenergy spinor (the superenergy spinor for the Riemann
(candidate) spinor) [2,4] and the Lanczos superenergy spinor (the superenergy spinor for the Lanczos (candidate) spinor), [4],
do not
have such a simple structure as can be seen below:
$$\eqalign{ {\cal B}_{ABCDA'B'C'D'} & = 4(\hat \chi_{ABCD}\bar{\hat\chi}_{A'B'C'D'} 
+\hat
\Phi_{ABC'D'}\bar{\hat {\Phi}}_{CDA'B'} ) \cr & =   4 \Bigl(\hat
\Psi_{ABCD}+
\hat\Lambda(\epsilon _{AC}\epsilon _{BD} +\epsilon _{AD}\epsilon _{BC})\Bigr)\cr
&\qquad\quad  \times \Bigl(\bar{\hat {\Psi}}_{A'B'C'D'}+\hat\Lambda(\epsilon _{A'C'}\epsilon _{B'D'}+\epsilon
_{A'D'}\epsilon _{B'C'}) \Bigr)+4\hat \Phi_{ABC'D'}\bar{\hat
{\Phi}}_{CDA'B'}\cr 
& =   4\Bigr(\hat
\Psi_{ABCD}\bar{\hat {\Psi}}_{A'B'C'D'} + \hat \Phi_{ABC'D'}\bar{\hat
{\Phi}}_{CDA'B'}  \cr   & \qquad \quad + \hat\Lambda^2 (\epsilon _{AC}\epsilon _{BD} +\epsilon _{AD}\epsilon _{BC})(\epsilon
_{A'C'}\epsilon _{B'D'}+\epsilon _{A'D'}\epsilon _{B'C'}) \cr & \qquad \qquad   + \hat\Lambda \hat
\Psi_{ABCD}(\epsilon _{A'C'}\epsilon _{B'D'}+\epsilon
_{A'D'}\epsilon _{B'C'})
+\hat\Lambda\bar{\hat {\Psi}}_{A'B'C'D'}(\epsilon _{AC}\epsilon _{BD} +\epsilon _{AD}\epsilon _{BC}) 
\Bigr)
}\eqno(13)$$
(where we have used the notation ${\cal B}$ rather than the more consistent ${\cal
T}[\hat R]$  simply for ease of presentation),
and 
$${\cal T}[\hat L]_{ABCDA'B'C'D'} = 2\Bigl(\hat L_{ABCD'}\bar {\hat L}_{A'B'C'D}+\hat L_{ABDC'}\bar {\hat
L}_{A'B'D'C}\Bigr)
\eqno(14)$$
where
$\hat L_{ABCD'}= \hat L_{(ABC)D'}
$.

It is easy to see that  simple identities such as (12) do not hold in these
cases. However this does not rule out the possibility of other, more complicated, identities. We shall look further at both
of these tensors in Section 6, and at possible identities for the Bel tensor in Section 4.

\

\

{\bf 3. Simple Tensor Identities.}

We now wish to confirm the tensor versions of  the four identities in Theorem 2 by tensor
means. We shall discover that although the above four identities had essentially the same spinor proofs, the
proofs for their superenergy tensor counterparts require very different amounts of calculations. We first give
the corresponding
$n$-dimensional basic superenergy tensors of the appropriate double $2$-forms.

In  $n$-dimensional spaces from [3], 

$\bullet$ the basic Bel-Robinson tensor [1] (equivalent to the Bel-Robinson spinor ${\cal
T}[\hat\Psi]$) is given by
$$ {\cal T}[\hat C]_{abcd}= \hat C_{apcq}\hat C_{b}{}^p{}_d{}^q
+\hat C_{apdq}\hat C_{b}{}^p{}_c{}^q- {1\over 2} g_{ab}
\hat C_{rpcq}\hat C^{rp}{}_d{}^q - {1\over 2} g_{cd}
\hat C_{aprq}\hat C_b{}^{p}{}^{rq}+{1\over 8} g_{ab} g_{cd}
\hat C_{sprq}\hat C^{sp}{}^{rq} \ ,
\eqno(15)$$
$\bullet$ the basic  trace-free Ricci superenergy tensor  (equivalent to the superenergy spinor for
the Ricci (candidate) spinor ${\cal T}[\hat\Phi]$) is given --- via the tensor\footnote {${}^{\dag}$}{Note that the
superenergy tensor constructed for the trace-free Ricci candidate tensor $\hat S_{ab}$ via the double 2-form $\hat E_{abcd}$ is
different from the superenergy tensor constructed for the trace-free Ricci candidate tensor directly via the double 1-form
${\hat S_{ab}}$ [3].} 
$\hat E_{abcd}= (\hat S_{ac}g_{bd}-\hat S_{ad}g_{bc}+\hat S_{bd}g_{ac}-\hat S_{bc}g_{ad}
)/(n-2)$
--- by, 
$$\eqalign{{\cal T}[\hat E]_{abcd} & = 
\hat E_{aecf} \hat E_b{}^e{}_d{}^f + \hat E_{aedf} \hat E_b{}^e{}_c{}^f -{1\over 2}
g_{ab} \hat E_{efcg}\hat E^{ef}{}_d{}^g -{1\over 2}
g_{cd} \hat E_{aefg}\hat E_b{}^{efg}+ {1\over 8}
g_{ab}g_{cd} \hat E_{efgh}\hat E_{efgh} \cr
& = 4\Bigl( \hat S_{ab}
\hat S_{cd}+ {n-4\over 2} \hat S_{a(c}\hat S_{d)b}- \hat S_{bp}\hat S_{(d}{}^p
g_{c)a}- \hat S_{ap}\hat S_{(d}{}^p
g_{c)b}+ {6-n \over 4}
\hat S_{cp}\hat S_d{}^p g_{ab}+ {6-n \over 4}
\hat S_{ap}\hat S_b{}^p g_{cd}\cr
& \qquad  + {n-6\over 8}
 g_{ab}g_{cd}\hat S_{pq}\hat S^{pq} + {1\over 2}
 g_{a(c}g_{d)b}\hat S_{pq}\hat S^{pq}\Bigr)/(n-2)^2
\ ,} \eqno(16)$$

$\bullet$ the basic  Ricci  scalar superenergy tensor    (equivalent to the superenergy spinor for
the Ricci (candidate) scalar $ {\cal T}[\hat\Lambda]$) is given --- via the tensor $\hat G_{abcd}=
\hat R(g_{ac}g_{bd}- g_{ad}g_{bc})/n(n-1)$ --- by
$$\eqalign{{\cal T}[\hat \Lambda]_{abcd} & = 
\hat G_{aecf} \hat G_b{}^e{}_d{}^f + \hat G_{aedf} \hat G_b{}^e{}_c{}^f -{1\over 2}
g_{ab} \hat G_{efcg}\hat G^{ef}{}_d{}^g -{1\over 2}
g_{cd} \hat G_{aefg}\hat G_b{}^{efg}+ {1\over 8}
g_{ab}g_{cd} \hat G_{efgh}\hat G^{efgh}\cr
& = {\hat R}^2\Bigl(2(n-2)g_{a(c}g_{d)b}+{n^2-9n+16 \over 4}g_{ab}g_{cd}\Bigr)/n^2(n-1)^2
\  ,} \eqno(17)$$

$\bullet$ the basic superenergy tensor for the  $\hat \chi$ (candidate) tensor  (equivalent to the
superenergy spinor for the $\hat \chi$ (candidate) spinor $ {\cal T}[\hat\chi]$) is
given --- via the tensor 
$\hat
\chi_{abcd}  = \hat C_{abcd}+
\hat R(g_{ac}g_{bd}- g_{ad}g_{bc})/n(n-1)$ --- by
$$\eqalign{{\cal T}[\hat \chi]_{abcd} & =
\hat \chi_{aecf} \hat \chi_b{}^e{}_d{}^f + \hat \chi_{aedf} \hat \chi_b{}^e{}_c{}^f -{1\over 2}
g_{ab} \hat \chi_{efcg}\hat \chi^{ef}{}_d{}^g -{1\over 2}
g_{cd} \hat \chi_{aefg}\hat \chi_b{}^{efg}+ {1\over 8}
g_{ab}g_{cd} \hat \chi_{efgh}\hat \chi^{efgh}\cr & 
= {\cal T}[\hat C]_{abcd}  +{\cal T}[\hat \Lambda]_{abcd} +2\hat R\bigl(\hat C_{acbd}+ \hat C_{adbc}\bigr)/n(n-1)
\  .} \eqno(18)$$

In $4$-dimensional space [3, 26],

$\bullet$ ${\cal T}[\hat C]_{abcd}$   has the same form as the
$n$-dimensional case

$\bullet$  ${\cal T}[\hat E]_{abcd}$ simplifies to     
$${\cal T}[\hat E]_{abcd} =
\hat S_{ab}
\hat S_{cd}+ \hat S_{ap}\hat S_b{}^p g_{cd}+ \hat S_{cp}\hat S_d{}^p g_{ab}- 3 \hat S_{p(a}\hat S_b{}^p g_{cd)}+{1\over 4}
\hat S_{pq}\hat S^{pq} (g_{ac}g_{bd}+g_{ad}g_{bc}-g_{ab}g_{cd}) \ ,
\eqno(19)$$ 

$\bullet$  ${\cal
T}[\hat \Lambda]_{abcd}$ simplifies to     
$${\cal T}[\hat \Lambda]_{abcd}=  {\hat R}^2(4g_{a(c}g_{d)b}-g_{ab}g_{cd})/144 \ ,
\eqno(20)$$
$\bullet$ ${\cal
T}[\hat \chi]_{abcd}$ simplifies  to
$$\eqalign{{\cal T}[\hat \chi]_{abcd}   
= {\cal T}[\hat C]_{abcd} +{\hat R\over 6} (\hat C_{acbd}+ \hat C_{adbc}) +{\cal T}[\hat \Lambda]_{abcd}
\  .} \eqno(21)$$

It is clear that all of the above superenergy tensors have the properties
$${\cal T}_{abcd}={\cal T}_{(ab)(cd)} \eqno(22)
$$
and  some have additional symmetry properties, e.g., for the Bel tensor (in four and five
 dimensions), and  the Lanczos superenergy tensor (in four dimensions),  
${\cal T}_{abcd}={\cal T}_{(abcd)} 
$ as we shall
show in Section 6. All of the above are labelled {\it basic} superenergy tensors to distinguish from the more {\it general}
superenergy tensors which can be obtained by taking linear combinations --- with positive constant coefficients
---  of different basic superenergy tensors, obtained by index permutations [3]. 

We now give the $4$-dimensional tensor counterparts of Theorem 2.

{\bf Theorem 2a.} 
In  $4$-dimensional spaces
the Bel-Robinson tensor 
${\cal T}[C]_{abcd}$ in (15)
satisfies 
$$
{\cal T}[\hat C]_{abcx} {\cal T}[\hat C]^{abcy}  = \delta^y_x \ {\cal T}[\hat C]_{abcd} {\cal
T}[\hat C]^{abcd}/4
\eqno(23)$$
 {\bf Proof.}  Substituting directly we obtain
$$\eqalign{{\cal T}&[\hat C]_{abcx}{\cal T}[\hat C]^{abcy} - {1\over 4} \delta^y_x \ {\cal
T}[\hat C]_{abcd}{\cal T}[\hat C]^{abcd}  \cr &   = 2 \hat C^{yb}{}_{cd}\hat C^{de}{}_{bf}\hat C^{fg}{}_{eh} \hat
C^{hc}{}_{gx} + 2
\hat
C^{yb}{}_{cd}\hat C^{de}{}_{bf}\hat C^{fgc}{}_{h} \hat C^{h}{}_{egx}   -
2 \hat C^{ab}{}_{cd}\hat C^{cd}{}_{eb}\hat C^{e}{}_{gh}{}^y \hat C_a{}^{gh}{}_{y} \cr & \qquad  -
 \hat C^{yb}{}_{cd}\hat C^{cd}{}_{eb}\hat C^{ef}{}_{gh} \hat C^{gh}{}_{xf}  +
\hat C_{abcd}\hat C^{abcd}C^{ey}{}_{gh}\hat C^{gh}{}_{ex}  \cr & \qquad \qquad  + {1\over 4}\delta_x^y \hat C^{ab}{}_{cd}\hat
C^{cd}{}_{eb}\hat C^{ef}{}_{gh} \hat C^{gh}{}_{af} -{1\over 16}\delta_x^y \hat C^{abcd}\hat C_{abcd}\hat
C^{efgh} \hat C_{efgh} \cr & \qquad \quad \qquad -{1\over 4} \delta^y_x\Bigl(2\hat C^{ab}{}_{cd}\hat
C^{de}{}_{bf}\hat C^{fg}{}_{eh}
\hat C^{hc}{}_{ga} + 2
\hat
C^{ab}{}_{cd}\hat C^{de}{}_{bf}\hat C^{fgc}{}_{h} \hat C^{h}{}_{ega}   \cr & \quad\qquad \qquad\qquad \qquad - 2
\hat C^{ab}{}_{cd}\hat C^{cd}{}_{eb}\hat C^{ef}{}_{gh} \hat C^{gh}{}_{af} 
+ {1\over 16} \hat C^{abcd}\hat
C_{abcd}\hat C^{efgh} \hat C_{efgh}
\Bigr)\ .  }
\eqno(24)$$
Using Lemma 4a a number of times gives the simpler expression
$$
\eqalign{{\cal T}[\hat C]_{abcx}{\cal T}[\hat C]^{abcy} - {1\over 4} &\delta^y_x \ {\cal
T}[\hat C]_{abcd}{\cal T}[\hat C]^{abcd} \cr & = 2 \hat C^{yb}{}_{cd}\hat C^{de}{}_{bf}\hat C^{fg}{}_{eh} \hat
C^{hc}{}_{gx} + 2
\hat
C^{yb}{}_{cd}\hat C^{de}{}_{bf}\hat C^{fgc}{}_{h} \hat C^{h}{}_{egx}  \cr & \quad  \quad - {1\over 4}
\delta^y_x\Bigl(2\hat C^{ab}{}_{cd}\hat C^{de}{}_{bf}\hat C^{fg}{}_{eh}
\hat C^{hc}{}_{ga}+ 2
\hat
C^{ab}{}_{cd}\hat C^{de}{}_{bf}\hat C^{fgc}{}_{h} \hat C^{h}{}_{ega} 
\Bigr)\ .  }\eqno(25)$$
We can now apply Lemma 4b to the first and third  terms, and Lemma 4c to the second and fourth terms, to obtain
the required result. \hfill $\diamondsuit$

\smallskip

{\bf Theorem 2b.}

 In  $4$-dimensional spaces, 
the  superenergy  tensor 
${\cal T}[\hat E]_{abcd}$ given in (19) 
satisfies 
$$
{\cal T}[\hat E]_{abcx}{\cal T}[\hat E]^{abcy}  = \delta^y_x \ {\cal T}[\hat E]_{abcd} {\cal T}[\hat E]^{abcd}/4
\eqno(26)$$

{\bf Proof.} 

 Substituting directly we obtain  
$$\eqalign{{\cal T}&[\hat E]_{abcx}{\cal T}[\hat E]^{abcy} - {1\over 4}  \delta^y_x \ {\cal T}[\hat
E]_{abcd}{\cal T}[\hat E]^{abcd}  \cr & = -  3 \hat S^y{}_b  \hat S^b{}_c
\hat S^c{}_d
\hat S^d{}_x  + {3\over 2} \hat S^{yb} \hat S_{xb} \hat S_{cd} \hat
S^{cd} + \hat S^y{}_x \hat S^a{}_b \hat S^b{}_c \hat
S^c{}_a   + {3\over 4}\delta^y_x \hat S^a{}_b \hat S^b{}_c \hat S^c{}_d \hat
S^d{}_a - {3\over 8} \delta^y_x \hat S^{ab} \hat S_{ab} \hat S_{cd} \hat
S^{cd}
\ .}  \eqno(27)$$
But the right-hand side of this equation is precisely 
$$ \delta^y_{[x} \ \hat S^a{}_{a} \hat S^b{}_{b}
\hat S^c{}_c
\hat S^d{}_{d]}\equiv 0 ,
\eqno(28)$$
where we have made use of Lemma 1. \hfill $\diamondsuit$

\smallskip

{\bf Theorem 2c.}

 In  $4$-dimensional spaces, 
the  superenergy  tensor 
${\cal T}[\hat \Lambda]_{abcd}$ given in (20) 
satisfies 
$$
{\cal T}[\hat \Lambda]_{abcx} {\cal T}[\hat \Lambda]^{abcy}  = \delta^y_x \ {\cal T}[\hat \Lambda]_{abcd} {\cal
T}[\hat \Lambda]^{abcd}/4
\eqno(29)$$

{\bf Proof.} The result follows from a direct calculation. \hfill $\diamondsuit$

\smallskip

{\bf Theorem 2d.}

 In  $4$-dimensional spaces,
the  superenergy  tensor 
${\cal T}[\hat \chi]_{abcd}$  given in (21) 
satisfies 
$$
{\cal T}[\hat \chi]_{abcx} {\cal T}[\hat \chi]^{abcy}  = \delta^y_x \ {\cal T}[\hat \chi]_{abcd} {\cal T}[\hat
\chi]^{abcd}/4
\eqno(30)$$

{\bf Proof.} 
$$
\eqalign{{\cal T}&[\hat \chi]_{abcx}  {\cal T}[\hat \chi]^{abcy}   \cr & = \Bigl({\cal T}[\hat C]_{abcx}+{\cal
T}[\hat
\Lambda]_{abcx}+ {\hat R\over 6} (\hat C_{acbx}+ \hat C_{axbc})\Bigr)
\Bigl({\cal T}[\hat
C]^{abcy}+{\cal
T}[\hat
\Lambda]^{abcy}  +{\hat R\over 6} (\hat C^{acby}+ \hat C^{aybc}) \Bigr)  \cr &
={\cal T}[\hat C]_{abcx}{\cal T}[\hat
C]^{abcy}+{\cal
T}[\hat
\Lambda]_{abcx}{\cal
T}[\hat
\Lambda]^{abcy} + {\hat R \over 6}\Bigl( {\cal T}[\hat
C]^{abcy}(\hat C_{acbx}+ \hat C_{axbc})+{\cal T}[\hat C]_{abcx}(\hat C^{acby}+ \hat C^{aybc}) \Bigr)  \cr &
\qquad +{{\hat R}^2\over 144}\hat C_{abcd} \hat C^{abcd} \delta_x^y }\eqno(31)$$
where the last term was obtained using Lemma 4a.

We can apply Theorems 2a and 2c to the first and second terms respectively; however, to complete the proof we
need to use {\it cubic} identities for the Weyl tensor; the details are given in [23]. 
\hfill $\diamondsuit$

\medskip We note that in Theorems 2,a,b,d explicit $4$-dimensional fddis were used in the proofs; it seems
unlikely that {\it direct} generalisations can be obtained by using analogous fddis in higher dimensions; more
likely, higher order identities would need to be considered. Theorem 2c was obtained by a direct calculation,
and in fact an analogous identity is clearly obtainable in $n$ dimensions because of the very simple structure
of the superenergy tensor in this case.

\

\

{\bf 4. Absence of simple identities for the Bel  Superenergy Tensor.}

The Bel tensor, the superenergy tensor for the Riemann tensor, in $n$ dimensions [3]
is
$${\cal B}_{abcd}= \hat R_{apcq}\hat R_{b}{}^p{}_d{}^q
+\hat R_{apdq}\hat R_{b}{}^p{}_c{}^q- {1\over 2} g_{ab}
\hat R_{rpcq}\hat R^{rp}{}_d{}^q - {1\over 2} g_{cd}
\hat R_{aprq}\hat R_b{}^{p}{}^{rq}+{1\over 8} g_{ab} g_{cd}
\hat R_{sprq}\hat R^{sp}{}^{rq}\eqno(32)
$$ 
with the obvious properties
$${\cal  B}_{abcd} = {\cal  B}_{(ab)cd} = {\cal  B}_{ab(cd)} = {\cal 
B}_{cdab}, \qquad {\cal  B}^a{}_{acd} = 0,
\eqno(33)$$
(We continue to  use the notation ${\cal B}_{abcd}$ rather than the more consistent ${\cal T}[\hat R]_{abcd}$.)

Substituting the usual decomposition of the Riemann tensor gives the alternative form [3], [15]
$${\cal  B}_{abcd} = {\cal T}[\hat C]_{abcd} + {\cal T}[\hat E]_{abcd}  + {\cal Q}_{abcd}\eqno(32')
$$
where 
$$\eqalign{{\cal Q}_{abcd}= & {1\over n-2} \Bigl( -4 \hat C^i{}_{(cd)(a} \hat S_{b)i} - 4 \hat C^i{}_{(ab)(c} \hat
S_{d)i} + 2 \hat S _{ij}\bigl(\hat C_a{}^j{}_{(c}{}^i g_{d)b}- \hat C_c{}^j{}_d{}^i g_{ab} +
\hat C_b{}^j{}_{(c}{}^ig_{d)a}- \hat C_a{}^j{}_b{}^i g_{cd}\bigl)\Bigr) \cr & \qquad + {2 \hat R \over n(n-1)}  (\hat
C_{acbd}+\hat C_{adbc}) \ .}\eqno(34)$$

In four dimensions we get 
$${\cal Q}_{abcd}= {\hat R\over 6} (\hat C_{acbd}+\hat
C_{adbc}) \ .$$

This last simplification is not obvious in tensors, although it is in spinors (11d); by tensors, it is obtained either  by
manipulation with duals [3], or via a
$4$-dimensional ddi, [15]
$$\hat S^e{}_f \hat C_{[ab}{}^{[cd} \delta_{e]}^{f]} \equiv 0 \ .\eqno(35)
$$

 The  Bel tensor is a generalisation of  the Bel-Robinson tensor, and an obvious
question is  whether it  
 also satisfies similar types of quadratic identity in four dimensions as the Bel-Robinson tensor
does. From spinor considerations it does not look very hopeful, so we investigate the possibility via  examples
rather than look for general results.

Because of the additional terms in the Bel tensor compared to the Bel-Robinson tensor we introduce
$$  
\qquad {\cal  B}_a{}^c{}_{cb} = {\cal  B}_{ab} = {\cal  B}_{ba}, \qquad {\cal  B} = 
{\cal  B}^a{}_a
\eqno(36)$$ and so we consider the  general   quadratic
identity with two free indices with the structure 
 $$k_1 {\cal  B}_{abcx}{\cal  B}^{abcy} + k_2 {\cal  B}_{abcx}{\cal 
B}^{acby} +k_3 {\cal  B}_{abx}{}^{y}{\cal  B}^{ab} + k_4 {\cal 
B}_{axb}{}^{y}{\cal  B}^{ab}+ k_5 {\cal  B}_{ax}{\cal  B}^{ay}+ k_6 {\cal 
B}{\cal  B}_x{}^{y} \ \propto \ \delta^y_x 
\eqno(37)$$
with constants $k_1,...,k_6$. By  substituting the Bel tensors of explicit spaces\footnote{${}^{\dag}$}{In
fact the choice of the van Stockum metric [24] $ds^2 = -dt^2-2a\rho^2 dt d\phi-e^{-2a\rho}d\rho^2
+e^{-2a\rho}dz^2 +(\rho^2-a^2
\rho^4)d\phi^2$ (available in {\it GRTensorII}  [25]) will give a one parameter solution for the constants used
above, and substituting these values leads to the identity (38).}  in the left hand side, we are led to
conjecture that, in 
$4$-dimensional spaces,  the  Bel tensor  satisfies the 
 quadratic identity 
$$2{\cal  B}_{a[bc]x}{\cal  B}^{a[bc]y}-{\cal  B}_{xa}{\cal  B}^{ay}+ 2{\cal 
B}_{ab}{\cal  B}_{a}{}^{[by]}{}_x + {1\over 2} {\cal  B} {\cal  B}_x{}^y =
\delta_x^y \ (2{\cal  B}_{a[bc]d}{\cal  B}^{a[bc]d}- 2{\cal  B}_{ab}{\cal 
B}^{ab} + {1\over 2} {\cal  B}^2)/4 \ .
\eqno(38)$$

So it appears that the Bel tensor {\it may} have a quadratic identity, which rather surprisingly does not reduce
to the Bel-Robinson identity in the vacuum case. However, closer inspection reveals that this identity is {\it
trivial} in the following sense:  the properties 
${\cal B}^{a}{}_{[bc]}{}^{d}={\cal B}^{[a}{}_{[bc]}{}^{d]}={\cal B}_{[b}{}^{[ad]}{}_{c]}, \ {\cal B}_{a[bcd]}=0$
 mean that we
can consider $B_{abcd}\equiv{\cal
B}{^{[c}{}_{[ab]}{}^{d]}}$ as a Riemann candidate and (38) becomes 
$$\tilde B_{abcx} \tilde B^{abcy} = \delta_x^y \ \tilde B_{abcd} \tilde B^{abcd}/4
\eqno(39)$$
where $\tilde B_{abcd}$ is  the trace-free part of $ B_{abcd}$, i.e., its Weyl candidate. But (39) is just
the identity in Lemma 4a which is a consequence of {\it only}   the trace-free $2$-form structure and the fact
that we are in $4$-dimensional space; {\it it has nothing to do with the superenergy structure of ${\cal
B}_{abcd}$ as a linear combination of products of Riemann candidates}.

Hence this identity is of no interest to us
in the context of superenergy tensors, and so,

{\bf Theorem 3.}
In  $4$-dimensional spaces, 
  the  Bel superenergy  tensor  (32)
${\cal B}_{abcd} $,  for a  Riemann (candidate) tensor $\hat R_{abcd}$,
 does not satisfy any non-trivial quadratic identity with
the structure (37).
\hfill
$\diamondsuit$

\

\

{\bf 5. Absence of simple identities for  Bel-Robinson Superenergy tensor in higher  dimensions.}

As noted above, it has
been found [3,15] that  the
Bel-Robinson tensor is completely symmetric in {\it five} (and lower) dimensions.  This raises the question as 
to
whether the above quadratic identity  for the Bel-Robinson tensor is also valid in five
dimensions; or, more generally, whether there exists {\it any} quadratic identity with
two free indices for the Bel-Robinson tensor in {\it five} dimensions.

Since in five dimensions the Bel-Robinson tensor is still
fully symmetric, but not  trace-free, the most general  quadratic
identity with two free indices which could exist would have to have the structure,
$$k_1 {\cal  T}_{abcx}{\cal  T}^{abcy}     +k_2
{\cal  T}_{abx}{}^{y}{\cal  T}^{ab}  + k_3
{\cal  T}_{ax}{\cal  T}^{ay}+ k_4 {\cal  T}\,  {\cal  T}_x^{y} \propto \
\delta_x^y\eqno(40)
$$
for constants $k_1, k_2, k_3 , k_{4}$, where
$${\cal  T}_{abcd} = {\cal  T}_{(abcd)},  
\qquad {\cal  T}_{abc}{}^c = {\cal T}_{ab} = {\cal
T}_{ba}, \qquad {\cal  T}_a{}^a = {\cal  T}  \ .\eqno(41)
$$  
To try to retrace the complicated tensor calculations of Theorem 2a, and try to replace the $4$-dimensional
fddis used there with higher dimensional fddis would be a very complicated procedure; so we try first 
to obtain a simple counterexample, and we easily obtain  the following negative result,

{\bf Theorem 4.} In  $5$-dimensional spaces, the Bel-Robinson 
tensor
${\cal T}_{abcd}$ does not satisfy any non-trivial quadratic identity of the form (40).

{\bf Proof.} Generalising the $4$-dimensional Kerr metric $g^K_{ab}$ to
five dimensions as $ds^2= g^K_{ab}dx^a dx^b + dx_5^2 $, we
can calculate  the Bel-Robinson tensor explicitly, and when we substitute it
into the left hand side of the above expression (40), we obtain
$$k_1 {\cal  T}_{abcx}{\cal  T}^{abcy}     +k_2
{\cal  T}_{abx}{}^{y}{\cal  T}^{ab}  + k_3
{\cal  T}_{ax}{\cal  T}^{ay}+ k_4 {\cal  T}\,  {\cal  T}_x^{y} = K \delta_x^y  + (J-K) \delta_x^5\delta_5^y
\eqno(42)$$
where 
$$K=144M^4 k_1/(x^2+a^2y^2)^6, \  J=-36M^4(x^6-15a^2 y^2 x^4+15a^4 y^4x^2-a^6y^6)^2
\bigl(k_1+k_2+k_3+k_4 \bigr)
$$ in Boyer-Lindquist coordinates. 
So clearly, in general, there are no choices of the constants $k_1, k_2, k_3 , k_{4}$ which will give us a
non-trivial  identity. \hfill$\diamondsuit$

Note that we have used  Bel-Robinson tensors constructed from  Weyl tensors and not the more
general candidates in this proof. This not only gives a stronger result than if candidates had been used, but
was obtained very simply using {\it GRTensorII} [25]. 

In spaces of dimension $n>5$ the Bel-Robinson tensor is no longer completely
symmetric and so to investigate the most general  possible   quadratic
identity with two free indices we would need to consider a much more complicated  form
than in (40).

\

\

{\bf 6.  Index symmetry of  Bel-Robinson and Lanczos superenergy tensors.}

In this section we will determine the kernel fddi for two results in four dimensions; we will then show by
considering the analogous higher dimensional fddis how, in one case there is a simple generalisation to five (and
only five) dimensions, while in the other there is no direct generalisation to higher dimensions. 

The Bel-Robinson spinor (11a) is trivially symmetric in all indices. On the other hand, the only obvious
symmetries from the Bel-Robinson tensor (15) are
${\cal T}_{abcd}={\cal T}_{(ab)cd}={\cal T}_{ab(cd)}={\cal T}_{cdab}$. To check if it is fully symmetric in {\it
all} indices we   examine
$$ {\cal T}[\hat C]_{a[bc]d}= {1\over 4} \hat C_{adef}\hat C_{bc}{}^{ef}
-\hat C_{eaf[b}\hat C_{c]}{}^e{}_d{}^f- 
\hat C_{fge[a}g_{d][b}\hat C_{c]}{}^{efg} +{1\over 8} g_{a[b} g_{c]d}
\hat C_{sprq}\hat C^{sp}{}^{rq} \ .
\eqno(43)$$
We know from spinors that it must be symmetric in all indices in (at least) four dimensions, 
so the structure of (43) invites comparison with the $4$-dimensional 
ddi for the Weyl tensor in Lemma 3 
contracted with another Weyl tensor, i.e.,
$$0 \equiv \hat C_{[ib}{}^{[kl}\delta_{c]}^{a]} \hat C_{kl}{}^{id} \ , \eqno(44)
$$
the right hand side of which when expanded coincides precisely with (43).

To determine whether the same result is valid in five dimensions, we consider  the analogous   {\it five}
dimensional fddi in Lemma 3,
and when we construct 
$$0 \equiv  \hat C_{[bc}{}^{[ad}\ \delta_{ij]}^{kl]} \hat C_{kl}{}^{ij} \ , \eqno(45)
$$
we find that its expanded right hand side also coincides precisely with (43).

So we have demonstrated   that the  fact that the Bel-Robinson tensor (15) is  fully symmetric in {\it five (and
lower) dimensions}\footnote{${}^{\dag}$}{The {\it five} dimensional ddi (45)
$$ {1\over 4} \hat C_{adef}\hat C_{bc}{}^{ef}
-\hat C_{eaf[b}\hat C_{c]}{}^e{}_d{}^f- 
\hat C_{fge[a}g_{d][b}\hat C_{c]}{}^{efg} +{1\over 8} g_{a[b} g_{c]d}
\hat C_{sprq}\hat C^{sp}{}^{rq} \equiv 0 
$$
which we have just exploited, is of course also valid in four dimensions; in four dimensions we can use Lemma 4a
on the penultimate term and obtain the  similar but simpler {\it four} dimensional ddi,  
$$ {1\over 4} \hat C_{adef}\hat C_{bc}{}^{ef}
-\hat C_{eaf[b}\hat C_{c]}{}^e{}_d{}^f 
 -{1\over 8} g_{a[b} g_{c]d}
\hat C_{sprq}\hat C^{sp}{}^{rq} \equiv 0 \ .
$$
which is just the identity (44). 
 Deser [6] has pointed out the significance of  this identity (44) in the
symmetry structure of the Bel-Robinson tensor in four dimensions; here we also see the significance of the
$5$-dimensional counterpart (45) in the
symmetry structure of the Bel-Robinson tensor in {\it five} dimensions.} can be seen as a simple consequence of
one
$5$-dimensional fddi.  (This result  was originally obtained for four and five dimensions seperately 
using duals in [3], and subsequently by the present  method in [15].)

For higher dimensions, from the
viewpoint of fddis, we  note that the next fddi $ \hat C_{[ab}{}^{[fg}\ \delta_{cde]}^{hij]}=0 
$ has too many indices to yield (43) by a   contraction with one Weyl tensor. 
However, it is easy to show conclusively that ${\cal T}[\hat C]_{abcd}$ is not symmetric in higher dimensions by taking the
double trace [3, 15].

\smallskip

The Lanczos superenergy spinor has the obvious symmetries ${\cal T}[\hat L]_{ABCDA'B'C'D'}={\cal T}[\hat
L]_{(AB)CD(A'B')C'D'}={\cal T}[\hat L]_{AB(CD)A'B'(C'D')}$ and the more general  Lanczos superenergy spinor

$$\tilde{\cal T}[\hat L]_{ABCDA'B'C'D'}=\bigl({\cal T}[\hat L]_{ABCDA'B'C'D'}+{\cal T}[\hat
L]_{CDABC'D'A'B'}\bigr)\eqno(46)$$ is clearly symmetric in all indices. 

  The basic Lanczos
superenergy tensor  (equivalent to the superenergy spinor for the Lanczos (candidate) spinor
$ {\cal T}[\hat L]_{ABCDA'B'C'D'}$) is given by [3, 15] in $n$ dimensions 
$${\cal T}[\hat L]_{abcd}=\hat {L}_{aic}\hat
{L}_{b}{}^i{}_{d}+\hat {L}_{aid}\hat {L}_{b}{}^i{}_{c}-{1\over
2}g_{ab}\hat {L}_{ijc}\hat {L}^{ij}{}_{d}- g_{cd}\hat {L}_{aij}\hat {L}_b{}^{ij}+
{1\over 4} g_{ab}g_{cd}
\hat {L}_{ijk}\hat {L}^{ijk} \ .
\eqno(47)$$ 
It has the obvious   properties ${\cal
T}[\hat L]_{abcd}={\cal T}[\hat L]_{(ab)(cd)}$,  but not the block symmetry ${\cal T}_{abcd}={\cal T}_{cdab} $. The more
general superenergy tensor 
$$\tilde{\cal T}[ \hat L]_{abcd}=\Bigl({\cal T}[\hat L]_{abcd}+{\cal T}[\hat L]_{cdab}\Bigr)/2 \ , \eqno(48)
$$ 
which is equivalent to (46), does not obviously appear to be completely symmetric, as we know it must be in four
dimensions at least. To determine if
$\tilde{\cal T}[ \hat L]_{abcd}$ is symmetric over {\it all} indices we examine
$$\tilde{\cal T}[\hat L]^{a}{}_{[bc]}{}^{d}={1\over 4}\hat {L}_{bce}\hat
{L}^{ade} -\hat {L}^{[a}{}_{e[b}\hat {L}_{c]}{}^{|e]d}-{1\over
2}\delta^{[a}_{[b}\hat {L}_{|ef|c]}\hat {L}^{|ef|d]}- \delta^{[a}_{[b}\hat {L}_{c]ef}\hat {L}^{d]ef}+
{1\over 4} \delta^{a}_{[b}\delta_{c]}^{d}
\hat {L}_{ijk}\hat {L}^{ijk} \  . \eqno(49)
$$ 
The structure of (49) (including two deltas) invites comparison with the $4$-dimensional 
ddi for the Lanczos tensor in Lemma 2 
contracted with another Lanczos tensor, i.e., 
$$0 \equiv \hat L _{[ab}{}^{[e}\delta_{cd]}^{fg]}  L^{cd}{}_g \eqno(50)
$$ 
the right hand side of which when expanded coincides precisely with (49).
 So we retrieve the result in [15],

{\bf Theorem 5.} A  Lanczos  superenergy tensor $\tilde{\cal T}[ \hat L]_{abcd}=\Bigl({\cal
T}[\hat L]_{abcd}+{\cal T}[\hat L]_{cdab}\Bigr)/2$ where ${\cal
T}[\hat L]_{abcd}$ is given by (47) is symmetric in all indices in four 
dimensions. \hfill
$\diamondsuit$

To determine whether the same proof is valid in five dimensions, we consider  the analogous   {\it five}
dimensional fddi in Lemma 2,
and we immediately see that there are too many free indices to yield  (49) by a contraction with one Lanczos
tensor.

\

\

{\bf 7. Tensor derivation of new electromagnetic conservation law.}

We now wish to look at a particular Lanczos candidate, 
$$L_{abc}=F_{ab;c}
\eqno(51)$$
where $F_{ab}$ is an electromagnetic field tensor which satisfies
the source-free Maxwell's equations 
$$F^{a}{}_{b;a}=0=F_{[ab;c]}
\eqno(52)$$ and so
the properties $L^{a}{}_{ba}=0=L_{[abc]}$ of a Lanczos candidate are
automatically satisfied. Hence we could choose the   tensor (47) with the above substitution (51)
and obtain
$${\cal T}[\nabla F]_{abcd}=F_{ai;c}
{F}_{b}{}^i{}_{;d}+F_{ai;d}F_{b}{}^i{}_{;c}-{1\over
2}g_{ab}F_{ij;c}F^{ij}{}_{;d}- g_{cd}F_{ai;j}F_b{}^{i;j}+
{1\over 4} g_{ab}g_{cd}
F_{ij;k}F^{ij;k}
\eqno(53) $$ 
 as a superenergy tensor for the electromagnetic
field.  In fact Senovilla [3] has shown that a tensor ${\cal C}_{abcd}$, suggested by
Chevreton [27] as an analogue for  the Bel-Robinson tensor in an electromagnetic field,
is  just a linear combination of two such superenergy tensors
$${\cal C}_{abcd}= \Bigl({ \cal T}[\nabla F]_{abcd}+{\cal T}[\nabla F]_{cdab}\Bigr)/2 \ .
\eqno(54)$$
This tensor was shown in [27] to have important properties {\it in flat
space}: in particular it is divergence-free, but  this property is not
valid, in general, in curved spaces. Recently Bergqvist, Eriksson and Senovilla [5] have used the spinor equivalent of the
Chevreton tensor and given two interesting properties {\it in curved space} for source-free
Einstein-Maxwell fields:

$\bullet$ the Chevreton tensor is fully symmetric

$\bullet$ the  trace of the Chevreton tensor is divergence free.

We now wish to consider the tensor versions of these results. 
The first of these properties is just a special case of the result
previously derived in tensors for Lanczos candidates in
$4$-dimensional spaces [15], and given at the end of Section 4.  The second property was deduced from the
spinor form of the divergence of
${\cal C}_{abcd}$ and Bergqvist, Eriksson and Senovilla [5] remark that the proof of this result is far
from obvious from the tensor point of view.  We shall now demonstrate that the result in [5]
can be obtained in a direct and straightforward manner --- until the 
complication  at the last stages where {\it two  fddis valid in four dimensions
} have to be used explicitly.  

{\bf Theorem 6.} In
four dimensions,
the non-zero  trace ${\cal C}_{ab}\equiv {\cal C}_{abc}{}^c$ of the Chevreton superenergy tensor ${\cal
C}_{abcd}$ is symmetric, trace-free and divergence-free, i.e., ${\cal C}_{a}{}^b{}_{;b} = 0$. 

{\bf  Proof.}

The non-zero trace of the Lanczos superenergy tensor is given by
$$  {\cal C}_{ab} \equiv {\cal C}_{abc}{}^c = -
L_{aef}L_b{}^{ef}+ {1\over 4} g_{ab}L_{cef}L^{cef}
\eqno(55)$$
and it is clear that it is symmetric and trace-free. The divergence is 
$${\cal C}_{a}{}^{b}{}_{;b}= -
L_{aef}L^{bef}{}_{;b}  -
L_{aef;b}L^{bef}  +{1\over 2} L_{cef;a}L^{cef}
\eqno(56)$$
Now substituting (52) and simplifying gives
$$\eqalign{2{\cal C}_{a}{}^{b}{}_{;b}  =  & -2
F_{ae;f}F^{be;f}{}_{b}  -2
F_{ae;fb}F^{be;f}  + F_{ce;fa}F^{ce;f}  
\cr = &  -2
F_{ae;f}F^{be}{}_{;b}{}^{f}  -2
F_{ae;f} \Bigl(R^f{}_b{}^e{}_i F^{bi}-R^f{}_iF^{ie}\Bigr)
-2F_{ae;bf}F^{be;f} -2
F^{be;f} \Bigl(R_{fba}{}^i F_{ie}+R_{fbe}{}^i F_{ai}\Bigr) 
 \cr  & \qquad
+ F_{ce;fa}F^{ce;f}
 \cr = &  -2
F_{ae;f}F^{be}{}_{;b}{}^{f}  -2
F_{ae;f} \Bigl(R^f{}_b{}^e{}_i F^{bi}-R^f{}_iF^{ie}\Bigr)
-3F_{[ae;b]f}F^{be;f}+F_{eb;af}F^{be;f}
 \cr &  \qquad -2
F^{be;f} \Bigl(R_{fba}{}^i F_{ie}+R_{fbe}{}^i F_{ai}\Bigr)  +
F_{ce;fa}F^{ce;f}
 \cr = &
 -2
F_{ae;f}F^{be}{}_{;b}{}^{f} -3F_{[ae;b]f}F^{be;f}  +
F_{eb;fa}F^{be;f}+ 2F^{be;f}
R_{afe}{}^i F_{ib}\cr  & \qquad -2
F_{ae;f} \Bigl(R^f{}_b{}^e{}_i F^{bi}-R^f{}_iF^{ie}\Bigr)
 -2
F^{be;f} \Bigl(R_{fba}{}^i F_{ie}+R_{fbe}{}^i F_{ai}\Bigr)   +
F_{ce;fa}F^{ce;f}}\eqno(57)$$

Using the source-free Maxwell's equations (52) and rearranging gives 
$$\eqalign{2{\cal C}_{a}{}^{b}{}_{;b} = {1\over 2}    
F_{ef;a} R^{ef}{}_{ib} F^{bi}+2 F_{ae;f}R^f{}_iF^{ie}
 +2
F^{be;f} R_{baf}{}^i F_{ie}-F^{be;f}R^i{}_{fbe} F_{ai}  }\eqno(58)$$
and decomposing the Riemann tensor in four dimensions gives
$$\eqalign{2{\cal C}_{a}{}^{b}{}_{;b} =     
{1\over 2}F_{ef;a} C^{ef}{}_{ib} F^{bi}&
 +2
F^{be;f} C_{baf}{}^i F_{ie}-F^{be;f}C^i{}_{fbe} F_{ai}\cr & +2 F_{af;e}S^f{}_iF^{ie}
+ 2F^{ie;f}S_{if}F^{ae}-F^{ie;f}S_{af}F^{ie} \ .}\eqno(59)$$

We have already noted that  the  $4$-dimensional fddi in Lemma 3, $C_{[ab}{}^{[cd}\delta_{f]}^{e]}\equiv  0$, when
contracted with
 $L_{de}{}^f$  gives the ddi  (4); a further contraction with $F^{cb}$
gives
$$
\eqalign{0 \equiv 
2F^{cb}L_{[b}{}^{de}C_{c]eda}-{1\over2} F^{cb}L^{de}{}_a C_{decb} & + 2F_a{}^b L^{def}
C_{bdef}
\cr & = 2F^{ie}L_{e}{}^{bf}C_{ifba}-{1\over2} F^{ib}L^{fe}{}_a C_{feib} - F_a{}^i L^{bef}
C_{ifbe} \ , } 
\eqno(60)$$
and  the substitution $L_{abc}=F_{ab;c}$ into (60) means  that the first three terms on the right hand side of (59)
disappear.  Next we use Einstein's equations  and equate the trace-free Ricci tensor
$S_{ab}$ to the usual expression for the electromagnetic energy tensor
$$S_a{}^b= T_{a}{}^{b}= F_{ae}F^{be} - \delta^b_aF_{cd}F^{cd}/4  .
\eqno(61)$$ 
Then the last three terms on the right hand side of (59) can be rearranged to give
$$\Bigl(F_{[b}{}^{b}F_{c}{}^{c}F_{d}{}^{d} F_{e}{}^{e}\delta_{a]}^f\Bigr){}_{;f} \equiv 0
\eqno(62)$$
since the expression inside the brackets is identically zero in four dimensions by
virtue of Lemma 1, which in this context is equivalent to the algebraic Rainich identity $T^a{}_c T^c{}_b = \delta^b_a\ T_{ij}
T^{ij}/4$ where the energy-momentum tensor $T^a{}_b = F^a{}_cF^c{}_b - \delta^b_a\ F^{ji}F_{ij}/4$, [15].
\hfill $\diamondsuit$

We note that, in the proof, $4$-dimensional
fddis have been used explicitly on two occasions ( (60) and (62) ), as well as a decomposition of the Riemann tensor in four
dimension. This is why the spinor proof appears much simpler, since the corresponding calculations in spinors just never
occur.  The use of the Einstein equations for the electromagnetic energy momentum tensor (61) is an
important component of the proof; the last three terms on the right hand side of (59) cannot be removed by other means, such
as a $4$-dimensional ddi which was the means  used to remove the first three terms. 

From the point of view of a direct generalisation of this method to higher dimensions, it is clear that the  higher
dimensional analogues, (e.g., the $5$-dimensional fddi in Lemma 3) would not be sufficient to reduce the first three terms of
(59) to zero; nor would the $5$-dimensional Cayley-Hamilton theorem be sufficient to reduce the last three terms of
(59) to zero. So it would appear that there is
no {\it simple and direct} generalisation of Theorem 6 in higher dimensions; but of course this does not rule out more involved
generalisations. The fact that the algebraic Rainich identity was used in this $4$-dimensional proof would suggest that higher
dimensional analogues would require   higher dimensional algebraic Rainich identities; in five dimensions this has been shown
to be a cubic identity in the energy-momentum tensor
$T^a{}_b$.

It may be of
interest to note that the result is actually true more generally for the trace of the  superenergy tensor ${\cal
T}[\nabla F]_{abcd}$.

\

\

{\bf 8. New symmetric  Bel-Robinson tensor generalisations in higher 
dimensions.}

We have noted in Section 6 that although ${\cal T}[\hat C]_{abcd}$ is symmetric in four and five
dimensions this result does not generalise to higher dimensions. However, the
$5$-dimensional fddi which established this result has a counterpart in other dimensions; so we now investigate
whether we can obtain analogous symmetry properties for some other superenergy tensors in higher dimensions.

Lovelock [14] has pointed out that the  $n$-dimensional counterpart of Lemma 3 is,

{\bf Lemma 5.} In $n = 2p$ dimensions,  the trace-free double $(p,p)$-form  $ V_{i_1i_2 ...
i_p}{}^{j_1j_2 ... j_p}=
V_{[i_1i_2 ... i_p]}{}^{[j_1j_2 ... j_p]}$ satisfies
$$ V_{[i_1i_2 ... i_p}{}^{[j_1j_2... j_p}\ \delta_{i_{p+1}]}^{j_{p+1}]}=0 \ .\eqno(63)$$

This specialises in {\it  six dimensions (and lower)} for a trace-free double $(3,3)$-form to
$$ V_{[abc}{}^{[efg}\ \delta_{d]}^{h]}=0 \ .\eqno(64)
$$

The more general results in [15] include 
  
{\bf Lemma 6.} In $n \le 2p+1 $ dimensions,  the trace-free double $(p,p)$-form  $ V_{i_1i_2 ...
i_p}{}^{j_1j_2 ... j_p}=
V_{[i_1i_2 ... i_p]}{}^{[j_1j_2 ... j_p]}$ satisfies
$$ V_{[i_1i_2 ... i_p}{}^{[j_1j_2... j_p}\ \delta_{i_{p+1}}^{j_{p+1}}\delta_{i_{p+2}]}^{j_{p+2}]}=0 \
.\eqno(65)$$
This specialises in {\it seven dimensions (and lower)} for a trace-free double $(3,3)$-form to
$$ V_{[abc}{}^{[fgh}\ \delta_{de]}^{ij]} \equiv 0 \ .\eqno(66)
$$
This  fddi (66) is the analogue of the fddi (45) used to establish symmetry of the Bel-Robinson
tensor in five and four dimensions. So we expect (66) to establish symmetry for some generalisation of the 
Bel-Robinson tensor such as a {\it trace-free} double $(3,3)$-form in {\it seven (and lower) dimensions}. 

Senovilla [3] has given a basic superenergy tensor for the double $(3,3)$-form  $ K_{abc}{}^{def} =
K_{[abc]}{}^{[def]}$ in $n$-dimensions as 
$$\eqalign{{\cal T}[K]_{abcd}=  \Bigl( K_{apqcrs}K_b{}^{pq}{}_d{}^{rs} 
+K_{apqdrs}K_b{}^{pq}{}_c{}^{rs} - & {1\over 3} g_{ab}
K_{pqrcst}K^{pqr}{}_d{}^{st} - {1\over 3} g_{cd}
K_{apqrst}K_b{}^{pqrst}\cr & +{1\over 18} g_{ab} g_{cd}
K_{pqrstu}K^{pqrstu} \Bigr)/4\ , }
\eqno(67)$$
To keep things simple, and maintain the analogy with the Weyl candidate $\hat C_{ab}{}^{cd}$, we assume also that
$ K_{abc}{}^{def}
$ is {\it trace-free and (block) symmetric}, i.e.,
$$K_{abc}{}^{dea} =0, \qquad K_{abc}{}^{def}= K^{def}{}_{abc} \ .\eqno(68)$$

So ${\cal T}[K]_{abcd}$ clearly has the  symmetry properties in $n$-dimensions,
$${\cal T}[K]_{abcd}={\cal T}[K]_{(ab)cd}={\cal T}[K]_{ab(cd)}={\cal T}[K ]_{cdab}
\eqno(69)$$  

To determine if ${\cal T}[K]_{abcd}$ is symmetric in {\it all }
indices we examine
$$\eqalign{{\cal T}[K]^{a}{}_{[bc]}{}^{d} =  \Bigl(K^{a}{}_{pq}{}^d{}_{rs}K_{[b}{}^{pq}{}_{c]}{}^{rs} 
+  & K^{apq}{}_{[c|rs|}K_{b]pq}{}^{drs} 
 - {1\over
3} \delta^{a}_{[b} K^{|pqr|}{}_{c]st}K_{pqr}{}^{dst} \cr &
- {1\over 3} \delta_{[c}^{d}
K^{a}{}^{pq}{}_{|rst|}K_{b]pq}{}^{rst} +{1\over 18} \delta^{a}_{[b} \delta_{c]}^{d}
K_{pqr}{}^{stu}K^{pqr}{}_{stu} \Bigr)/4 \ ,}
\eqno(70)$$
The structure of (70) (including two deltas) suggests that we exploit the {\it seven dimensional identity} (66)
and investigate 
$$0\equiv  K_{[efg}{}^{[hij}\ \delta_{bc]}^{ad]}K^{efg}{}_{hij} \ .\eqno(71)
$$
When we write out (71) as
$$ \eqalign{0\equiv  \Bigl({1\over 3}& K_{bcp}{}^{qrs}K^{adp}{}_{qrs}-  {1\over 2} 
K_{pqc}{}^{fgd}K^{pqa}{}_{fgb}+{1\over 2} K_{pqc}{}^{fga}K^{pqd}{}_{fgb}  - {1\over 6}  \bigl(
K_{pqi}{}^{fga}K^{pqi}{}_{fgb}\delta^d_c \cr & -  
K_{pqi}{}^{fga}K^{pqi}{}_{fgc}\delta^d_b  +
K_{pqi}{}^{fgd}K^{pqi}{}_{fgc}\delta^a_b - 
K_{pqi}{}^{fgd}K^{pqi}{}_{fgc}\delta^a_c \bigr)  
+ {1\over 18} K^{pqr}{}_{fgh}K_{pqr}{}^{fgh}\delta^{ad}_{bc}\Bigr)/4
}\eqno(72)$$
we easily see that the right hand side of (72) does not coincide with (70), because of an apparent discrepency in
the respective first terms. (Using the symmetry properties (68) enables us to match up all the other terms.)

However, if we consider 
$K_{abcdef}$ to satisfy a {\it first Bianchi-type identity}   
$$K_{ab[cdef]}=0\eqno(73)$$
 --- as well as being
trace-free and (block) symmetric --- then we find that the first term on the right hand side of (70) becomes
$$K^{a}{}_{pq}{}^d{}_{rs}K_{[b}{}^{pq}{}_{c]}{}^{rs} = - K^a{}_{pqrs}{}^d
K_{bcpqrs}= K^{ad}{}_{pqrs}K_{bc}{}^{pqrs}/3
\eqno(74)$$
and now it is easy to see that the right hand side of (72) coincides term by term with (70), and so we
have proved 

{\bf Theorem 7.} In seven (and less) dimensions a  trace-free symmetric double $(3,3)$-form $K_{abc}{}^{def}$
which also satisfies
$K_{ab[cdef]}=0$ has a superenergy tensor ${\cal T}[K]_{abcd}$ given by (67)
which is symmetric in all indices. 

\hfill $\diamondsuit$

Again, in analogy with the Bel-Robinson tensor in four dimensions we can see from a direct calculation on the
index pair $(ab)$ in  (67),  combined with Theorem 7, that

{\bf Corollary 7.1.}  In six  dimensions,  the  trace-free symmetric double $(3,3)$-form $K_{abc}{}^{def}$
which also satisfies
$K_{ab[cdef]}=0$, has a superenergy tensor ${\cal T}[K]_{abcd}$, given by (67) which is symmetric in all indices
and trace-free on all pairs of indices.

 \hfill $\diamondsuit$

Let us  consider $K_{abcdef}$ to be also  {\it divergence-free}, i.e., 
$$K^{a}{}_{bcdef;a}=0 \ .\eqno(75)$$ 
It is then straightforward  to repeat the type of calculation which has been done for the divergence-free Weyl
tensor in Ricci-flat spaces in $n$ dimensions, and show that

{\bf Corollary 7.2.} 
If in addition $K_{abc}{}^{def}$  also satisfies $K^{a}{}_{bcdef;a}=0$
then its superenergy tensor ${\cal T}[K]_{abcd}$ is  also divergence free in $n$ dimensions.,
i.e., 
$${\cal T}[K]^a{}_{bcd;a}=0 \ .
\eqno(76) $$
\hfill $\diamondsuit$

We have  given just this one application as a simple example to illustrate the power of the fddis in
generalising a result to a higher dimension. 
However, from this pattern, we would expect to generalise Theorem 7 to classes of 
$(p,p)$-forms in spaces of dimension
$n=2p+1$. Similarly it appears likely that  the result in Theorem 5 for  Lanczos candidates in four dimensions,
could be generalised to classes of $(p,q)$-forms in $n=p+q+1$ dimensions.

\

\

{\bf 9. Discussion.}

We have used ddis in  many situations in this paper, both to rederive existing results, in an efficient tensorial manner, and
to obtain new results.  The four quadratic identities, which we consider in Sections 2 and 3,  are all special cases of one
spinor result which can be established very easily in spinors. We were able to derive the corresponding tensor identities
in a manner independent of signature. However, the tensor versions are of varying degrees of difficulty: the familiar
Bel-Robinson identity in Theorem 2a requires considerable preliminary work to establish lemmas for the Weyl tensor, and
Theorem 2d requires Theorem 2a together with a number of other lemmas, one involving a 'mixed' identity. The tensor derivation
of the new conservation law for electromagnetism [5] given  in Section 7 also requires  'mixed' ddis. These tensor derivations
are very complicated, and one wonders how long it would have taken to even conjecture   the results in  Theorems 2d and 6 
without the  parallel spinor (or null vector) results; but it seems that there is no easier signature-free way in tensors, and
we should train ourselves to recognise such structures in tensors.    Of course, none of the  fddis which we are using in this
paper are   'new', but while ddis which involve only one tensor (such as quadratic identities for the Weyl tensor, or the
Cayley-Hamilton theorem for the Ricci tensor), are familiar,  on the otherhand, 'mixed' ddis such as (4),
(35), (60), (62) involving more than one tensor and/or derivatives are much less familiar;  one of the
purposes of this paper is to draw attention to such possibilities.

The amount of work required in these tensor calculations in four dimensions serves as a warning of the even more
complicated calculations which will be required to establish analogous results in higher dimensions; the
existence of fddis as 'signposts' will be invaluable.

\smallskip

It is clear from the above examples that the exploitation of dimensionally
dependent identities is a useful method, not only for confirming suspected
tensor identities, but also for establishing new and perhaps unexpected results. For instance, when we have  a
particular tensor expression, a study of its structure can suggest an overlap between some of its terms and the
terms in a dimensionally dependent identity, and so we have an opportunity of
exploiting the latter, and discovering hitherto unexpected tensor
relationships. (This is actually how the unexpected symmetry property of the  Lanczos
superenergy tensor was first recognised in [15].) Furthermore, once a new significant tensor relationship is
established in four dimensions and the kernel $4$-dimensional fddi identified, then the analogous fddis
in five dimensions and higher can be investigated with the hope of establishing new tensor relationships in
these higher dimensions. 

 Theorem 7 is an example of how  to exploit this approach. We can continue to look for higher dimensional analogues of
significant $4$-dimensional results: the identity (63) for  the trace-free double $(p,p)$-form  $ V_{i_1i_2 ...
i_p}{}^{j_1j_2 ... j_p}=
V_{[i_1i_2 ... i_p]}{}^{[j_1j_2 ... j_p]}$ in dimensions $n=2p$ leads to the quadratic identity [14]
$$ V_{x i_2 ... i_p}{}^{j_1j_2... j_p}V_{j_1j_2 ... j_p}{}^{y i_2 ...
i_p}=
\ \delta_{x}^{y} \  V_{i_1i_2 ...
i_p}{}^{j_1j_2 ... j_p} V_{j_1j_2 ... j_p}{}^{i_1i_2 ...
i_p}/2p.$$
The quadratic identity (1) for the Bel-Robinson tensor (i.e., the superenergy tensor for the Weyl tensor where  $p=2$) in
four dimensions motivates the question as to whether there exists an analogous quadratic identity for the superenergy tensor of
(the block symmetric part of) $ V_{i_1i_2 ... i_p}{}^{j_1j_2 ... j_p}$ in  dimension $n=2p$.   The stronger versions,
and the higher dimensional generalisations will be presented elsewhere. One criticism of this new generalised Bel-Robinson
tensor
${\cal T}[K]_{abcd}$ would be  the apparent lack of explicit connection of
$K_{abc}{}^{def}$ with physical fields; we shall demonstrate in a future paper that there are indeed important   
links with the gravitational field as described by the Weyl tensor $C_{abcd}$, and that we can construct
examples of the tensor $K_{abc}{}^{def}$ which inherit some of the properties of $C_{abcd}$.

\smallskip

The result in Theorem 7 has illustrated one possible way to exploit  analogous fddis  in higher dimensions: to generalise to
forms of higher rank. Another approach has been taken in the generalised Rainich problem [16], where  higher
dimensional  identities analogous to 4-dimensional quadratic identities were obtained as cubic and higher order
identities --- involving very long manipulations. It seems clear from the tensor derivation  of the new
conservation law in electromagnetic theory, that {\it direct} generalisation to higher dimensions are not
possible. However, we would expect some sort of generalisation exploiting the higher dimensional counterparts
of the kernel fddis used in four dimensions. 
With the generalised Rainich results  [16] as 'signposts' we would speculate there may be a generalisation of the new
conservation law in five dimensions for  superenergy tensors involving {\it cubic} terms of 
$T_{ab}$.

\smallskip

While obviously interesting and useful in themselves, the study of  these  identities is not an end in itself.
What is of interest is to find  identities which are sufficient as well as necessary conditions for a
factorisation result, and to be able to study a 'generalised Rainich-Maisner-Wheeler  problem'. Clearly then we will need
the most general versions of such identities; for instance Penrose [2] has given the most general spinor version of
the spinor identity (12) for the Bel-Robinson tensor --- a quadratic identity with {\it all} indices free --- and
shown that it is also sufficient to achieve the factorisation (11a). The ddis which we have been studying in
this paper will give us the basic structures to continue these investigations.

\medskip

Finally, we would also emphasise how simple it is to disprove conjectured identities by counterexample; {\it
GRTensorII}, [25]  is an invaluable tool for this, and  it can also be efficient in enabling us to distinguish
between trivial and non-trivial identities. We have  also benefited from the use of {\it Tensign} [28], making
it  possible to guarantee the accuracy in the extensive index manipilation required in some of the results.

\

\

{\bf Acknowledgements.}

We are grateful to  Jos\'{e} Senovilla for suggesting an investigation of the Bel-Robinson quadratic
identity by dimensionally dependent identities, for  
supplying the reference [21], and pointing out the
applicability of results on the Lanczos superenergy tensor to the Chevreton tensor, as well as for other comments and
suggestions. We also thank G\"oran Bergqvist, Ingemar Eriksson,  Anders H\"oglund  and Magnus
Herberthson for discussions and suggestions. B.E. wishes to thank  Vetenskapsr\aa det (the Swedish
Research Council) for financial support.

\

\

{\bf References}

[1]  L. Bel, {\it CR Acad. Sci. Paris}, {\bf 246}, {3105}, (1958);
{\it CR Acad. Sci. Paris}, {\bf 247}, {1094}, (1958); {\it CR Acad. Sci. Paris}, {\bf 248}, {1297}, (1959); 
 I. Robinson, unpublished King's College Lectures (1958);
 {\it Class. Quant Grav.}, {\bf 14}, {4331}, (1997)

[2] R. Penrose and W. Rindler, {\it Spinors and spacetime}, 2 vols.
(Cambridge Univ. Press, Cambridge, 1984, 1986).

[3]  J. M. M. Senovilla, 
      {\it Class. Quant Grav.},
     {\bf 17}, 2799 (2000); J. M. M. Senovilla,  {\it Gravitation and Relativity in
General}, eds. A. Molina, J. Mart\'{\i}n, E.Ruiz and F. Atrio
(World Scientific, 1999), (preprint gr-qc/9901019).

[4]  G. Bergqvist, {\it Commun. Math. Phys}, {\bf 207}, 467, (1999); 
{\it J. Math. Phys.},
   {\bf 39}, 2141, (1998).
 
 [5] G. Bergqvist, I. Eriksson, J. M. M. Senovilla, {\it New electromagnetic conservation laws.} 
  
preprint gr-qc/0303036

[6]   S. Deser,  in {\it Gravitation and Relativity in
General}, eds. A. Molina, J. Mart\'{\i}n, E. Ruiz and F. Atrio
(World Scientific, 1999), (preprint gr-qc/9901007);
S. Deser, J.S Franklin and D. Seminara, {\it Class. Quant Grav.}, {\bf 16}, 2815, (1999);
S. Deser and D. Seminara, {\it Phys. Rev. } D62, 084010 (2000).

[7] S. B. Edgar,  {\it Gen. Rel. Grav.}, {\bf 31}, 405, (1999).

[8] F. Andersson and S. B. Edgar, {\it Int. J. of Mod. Phys} D, {\bf 5}, 217, (1996). 

[9]  C. Lanczos,  {\it Rev. Mod. Phys.}, {\bf 34}, 379, (1962).

[10] R. Illge, 
    {\it Gen. Rel. Grav},
    {\bf 20}, 551, (1988).

 [11] S. B. Edgar, {\it Mod. Phys.
Lett.} A,
    {\bf 9}, 479, (1994); 
 S. B.  Edgar and A. H\"oglund, 
    {\it Proc. R. Soc. Lond.} A,
    {\bf 453}, 835, (1997).

[12] Dianyan Xu, 
    {\it Phys. Rev. D},
    {\bf 35}, 769, (1987).

[13] I. Jack  and L. Parker, 
    {\it J. Math. Phys.},
    {\bf 28}, 1137, (1987).

[14] D. Lovelock, {\it Proc. Camb. Phil. Soc.} {\bf 68}, 345, (1970).

[15] S. B.  Edgar and A. H\"oglund, {\it J. Math Phys}, {\bf 43},  659, (2002).

[16] G. Bergqvist and A. H\"oglund, {\it Class. Quant Grav.}, {\bf 19},  3341, (2002).
 
[17] G. Bergqvist and J. M. M. Senovilla,   {\it Class. Quant Grav.}, {\bf 18},  5299, (2001).
 
[18]  S. B.  Edgar and A. H\"oglund, {\it Gen. Rel. Grav},
    {\bf 32}, 2307. (2000).

[19] S. Deser in {\it Gravitation and Geometry}, eds. W. Rindler and A. Trautman
(Naples, Bibliopolis, 1987)

[20] P. Dolan and C. W. Kim,  {\it Proc. R. Soc. Lond.} A,
    {\bf 447}, 557, (1994).

[21] V. D. Zakharov, {\it Gravitational Waves in Einstein's Theory}, (New York: Halsted, Wiley 1973).

[22] R. Debever, {\it Bulletin de la Soci\'{e}t\'{e} de Belgique}, {\bf 247}, 112, (1958). 

[23] O. Wingbrant, { M. Sc. dissertation}, LiTH-MAT-Ex-2003-09, Department of Mathematics, Link\"opings
universitet, Link\"oping, Sweden.

[24] W. J. van Stockum, {\it Proc. R. Soc. Edinburgh}, {\bf A57}, 135, (1937).

[25] GRTensorII: Available at http://grtensor.org

[26] M. \'{A}. G. Bonilla and J. M. M. Senovilla,
{\it Gen. Rel. Grav}, {\bf 29}, {91}, (1997).

[27] M. Chevreton, {\it Nuovo Cimento}, {\bf 34}, {901}, (1964).

[28] Tensign, information at http://www.lysator.liu.se/$\tilde{\ }$andersh/tensign/

\end